\documentclass{PoS}

\title{The QCD Phase Diagram: Expectations and Challenges}

\ShortTitle{The QCD Phase Diagram: Expectations and Challenges}

\author{\speaker{Reinhard Stock}\\
        Physics Department,
        University of Frankfurt, Germany\\
        E-mail: \email{stock@ikf.uni-frankfurt.de}}

\abstract{A survey is given of recent QCD theory advances concerning the phase diagram, in particular the indications for a critical point and adjacent first order phase transition at high baryo-chemical potential, and the new ideas concerning a further phase at high $\mu_{B}$, the quarkyonic matter phase. The latter state might finally explain the hadro-chemical equilibrium freeze-out points from A+A collisions at energies below SPS energy. We review several event-by-event fluctuation signals that promise to shed a light on the existence of a critical point, and we discuss its possible reflection in recent lattice QCD calculations.}

\FullConference{5th International Workshop on Critical Point and Onset of Deconfinement - CPOD 2009,\\
		 June 08 - 12 2009\\
		 Brookhaven National Laboratory, Long Island, New York, USA}

\begin{document}

\section{Introduction: Motivation}

The phase diagram of strongly interacting matter represents the, perhaps, most challenging
open problem within the Standard Model of elementary interaction. Its most prominent
feature, the deconfinement transition line between hadrons and partons, has been first
addressed by R. Hagedorn~\cite{1}, well before the advent of QCD, in his studies of the
limiting temperature occuring in hadron-resonance matter. The resulting phase boundary, at
about $T=160-170 \: MeV$ (that concurs with an energy density of about $1 \: GeV/fm^3$),
was subsequently understood~\cite{2,3} as the location of the QCD hadron to parton
deconfinement transition. At such low temperature, and corresponding mean $Q^2$,
deconfinement can not be the consequence of QCD asymptotic freedom - the perturbative QCD
mechanism that was first envisaged \cite{4,5} - but falls deeply into the non-perturbative
QCD sector, as does the resulting confined hadron structure. In fact, non-perturbative QCD
theory on the lattice has, for two decades, postulated that the hadron-parton phase
transformation occurs in the vicinity of Hagedorn's limiting temperature \cite{6,7}, at
zero baryochemical potential. \\
\begin{figure}[hbt]
\begin{center}
\includegraphics[width=9cm]{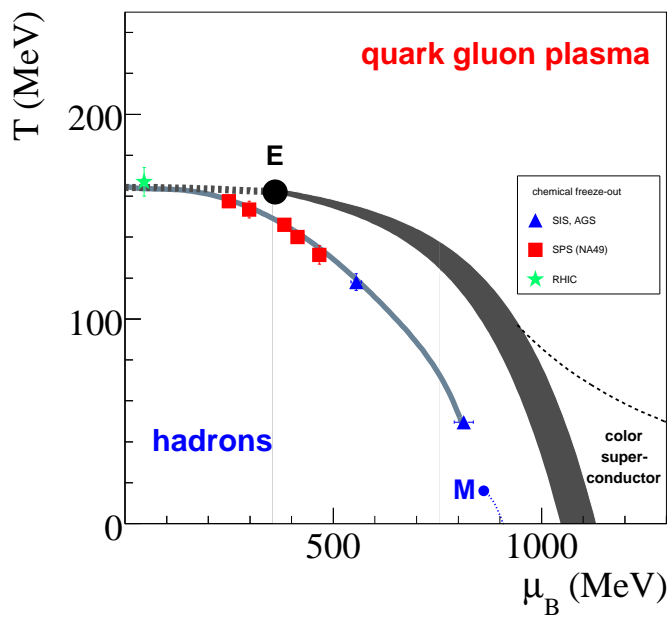}
\caption{Sketch of the QCD phase diagram in grand canonical variables $T, \: \mu_B$. A
critical point $E$ is indicated which ends the crossover transition domain at $\mu_B$
below $E$. Also shown are the hadronic chemical freeze-out points resulting from
statistical model analysis, from RHIC down to SIS energies.}
\label{fig:fig1}
\end{center}
\end{figure} \\
Conditions reached in the cosmological expansion evolution,
and closely approached at RHIC and LHC energies, in collisions of heavy nuclei.
Subsequent, recent developments in QCD Lattice theory have overcome the technical
limitation to the case of zero baryochemical potential \cite{8,9,10}, with extrapolations
of the deconfinement phase boundary, upward to about $\mu_B=500 \: MeV$ in the $(T,
\mu_B)$ plane, including hints of a critical point of QCD. This domain coincides with the
energy region of $A+A$ collision study at CERN SPS energies, $5<\sqrt{s}<17 \:GeV$, which
has created several non-trivial, much discussed experimental observations to which we
shall turn below. The expectations raised both, by the recent lattice QCD developments,
and by the former SPS experiments, concerning the physics of the phase boundary at
non-zero baryochemical potential, have recently been reflected in new experimental
efforts, notably with the plans for RHIC running at such low energy, and CERN re-running
with SPS experiment NA61, and with the FAIR facility developed at GSI. The present article
will deal with the corresponding physics goals, and challenges. We begin with a short
recapitulation. \\
\begin{figure}
\begin{center}
\includegraphics[width=6cm]{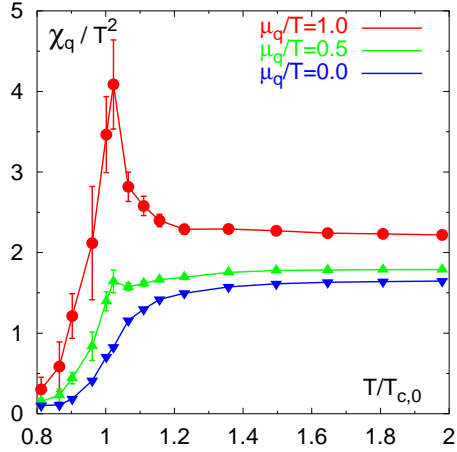}
\caption{Light quark number density fluctuation (susceptibility) from a two-flavour
lattice calculation \cite{25} for quark chemical potentials $\mu_q=0.75$ and $150 \: MeV$,
respectively.}
\label{fig:fig2}
\end{center}
\end{figure}\\
Our present view of the QCD phase diagram is sketched in Fig.~\ref{fig:fig1}. It is dominated
by the confinement boundary line above which some kind of parton plasma should exist. Up
to about $\mu_B=500 \: MeV$ it results from recent lattice QCD extrapolations
\cite{8,9,10}; its further extension stems from ''QCD-inspired'' theoretical
considerations \cite{11,12,13}. Lattice studies agree \cite{10,14} that the domain toward
$\mu_B=0$ should feature a rapid crossover transition, with a coincidence of deconfinement
and QCD chiral symmetry restoration. As the deconfinement transition at high $\mu_B$ is
generally expected \cite{11,12,13} to be of first order one is led to conclude on the
existence of a critical point $E$, at intermediate $\mu_B$. In fact lattice QCD has
reported first evidence for the existence of a critical point \cite{8,9,10,15}, to which
we shall turn below. Fig.~\ref{fig:fig1} also shows experimental information concerning the $(\mu_B,
T)$ values at which the hadron-resonance species population freezes out from the dynamical
evolution, in $A+A$ collisions of mass $A \approx 200$ nuclei at various energies ranging
from top RHIC, via SPS and AGS, down to SIS energy. Remarkably, this freeze-out occurs in
''hadro-chemical'' species population equilibrium, as reflected in the Grand Canonical
Gibbs ensemble of the statistical hadronization model \cite{16,17,18}. We observe that the
freeze-out points merge with the lattice QCD phase boundary, from top SPS to top RHIC
energies. It thus appears that hadron-resonance species freeze-out coincides with the QDC
confinement conditions, leading to hadronization, at low $\mu_B$. In fact, the idea that
the phase transformation ''gives birth'' to an equilibrium hadron-resonance population was
first formulated by Hagedorn \cite{19} and has been further explored recently
\cite{20,21,22,23}. We would, thus, locate experimentally the parton-hadron phase boundary
at $\mu_B \rightarrow0$. However, Fig.~\ref{fig:fig1} shows that the freeze-out points fall well below the conjectured confinement boundary as $\mu_B \rightarrow 500 \: MeV$. It remains an open, challenging question to identify a transport mechanism, active during expansion after confinement, which could deliver the system in species equilibrium at much lower temperatures. We turn to this topic in the next section but note, for now, that the recent hypothesis \cite{24} of a further QCD phase (labeled ''Quarkyonium''), occuring in between confinement and hadronization, might provide for another phase transition, at high
$\mu_B$, which enforces hadro-chemical equilibrium conditions at freeze-out.\\
\begin{figure}
\begin{center}
\includegraphics[width=7cm]{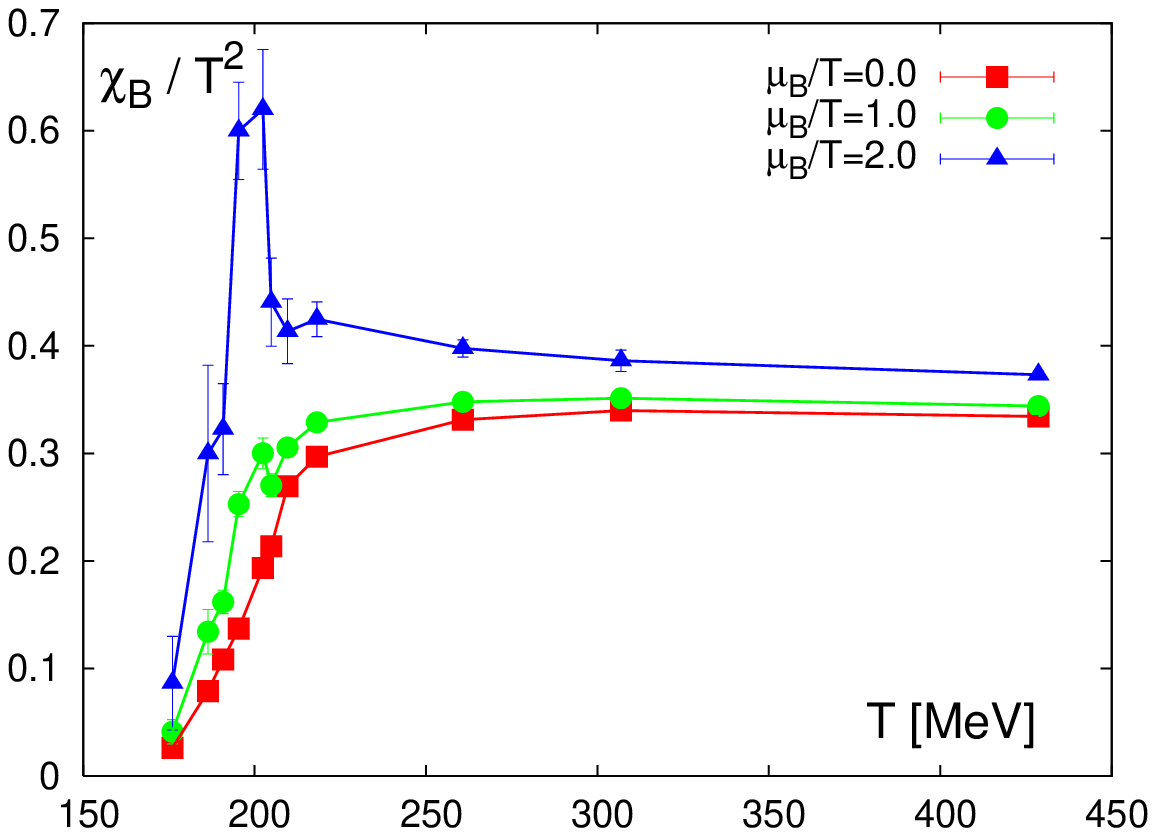}
\includegraphics[width=7cm]{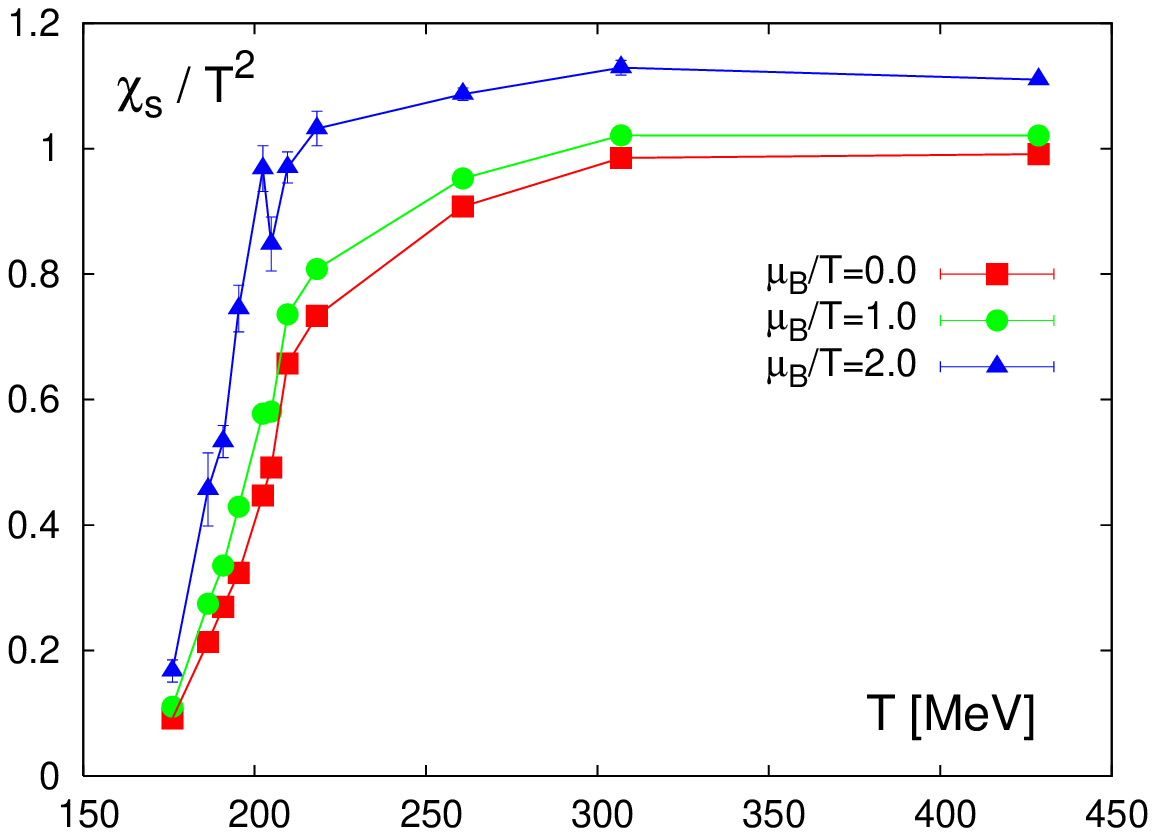}
\caption{Light and strange quark susceptibility from a 3-flavour lattice calculation
\cite{27} at baryochemical potential $\mu_B=(3\mu_q)= 0, \:T_c$ and $2 \: T_c$.}
\label{fig:fig3}
\end{center}
\end{figure}\\
Early consideration of a critical point at finite $\mu_B$ arose from quark
susceptibilities in lattice QCD \cite{25}, as illustrated in Fig.~\ref{fig:fig2}. The calculation
employed two dynamical quark flavours and showed a smooth (''crossover'') behaviour at the
critical temperature $T_c$ for $\mu_B=3\: \mu_q=0$, but a dramatic susceptibility peak
developing toward $\mu_B=3 \: T_C=450 \: MeV$. Such a quark number density fluctuation was
implied, as a characteristic consequence of a critical point, in the pioneering 1999
article of Stephanov, Shuryak and Rajagopal \cite{26}. The most recent lattice results, of
a 3 flavour calculation with near-realistic quark masses \cite{27} are shown in Fig.~\ref{fig:fig3}. The sharp peak structure in the light quark sector remains, at $\mu_B=2 \: T_c$, and even the strange quark susceptibility exhibits a (dampened) irregularity. 

\begin{figure}
\begin{center}
\includegraphics[width=8cm]{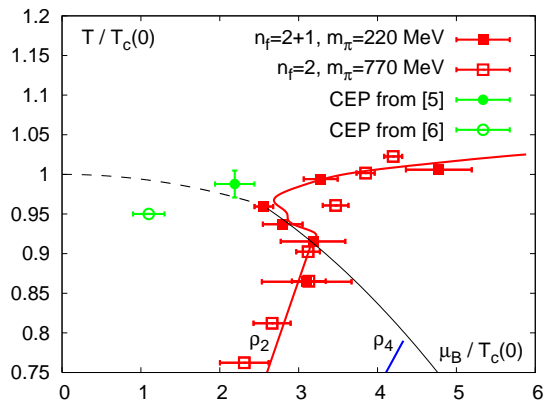}
\caption{Radius of convergence of the Taylor expansion coefficient for the pressure, in
the $T, \: \mu_B$ plane, from a lattice 3-flavour calculation \cite{28}, indicating non
monotonic structure at $\mu_B \approx 500 \: MeV$.}
\label{fig:fig4}
\end{center}
\end{figure}

\noindent Further toward gaining hints for the existence of a critical QCD point, from
state-of-the-art lattice calculations, C. Schmidt of the RBC-Bielefeld collaboration
showed \cite{28} at QM08 an analysis of the radius of convergence, observed in the Taylor
expansion coefficients (toward high $\mu_B$) of the pressure, as illustrated in Fig.~\ref{fig:fig4}. By definition, the Taylor expansion convergence ends at a critical point. In fact, the second order estimate of the radius of convergence exhibits a first indication of a nonmonotonic decrease at $T \approx 195 \: MeV$, $\mu_B \approx 500 \: MeV$ in this calculation. Higher orders in the pressure expansion are clearly required to pin down this first result - with obvious implications concerning lattice QCD computing power advances.

\noindent The energy domain, of interest to the search for manifestations of a critical
point, and the adjacent onset of a first order nature of the phase transition, is broadly
outlined by such lattice observations: to fall into the interval $250<\mu_B<500 \: MeV$.
This corresponds, almost exactly, to the SPS range $6.5<\sqrt{s}<17.3 \: GeV$, also
including the top energy reached in former AGS investigations. At the low end of this
interval, we might expect the ''onset of deconfinement'' \cite{29}: the dynamical
trajectory of central heavy nucleus collisions would reach its turning point (between
initial compression/heating and final expansion/cooling) in the close vicinity of the
parton-hadron boundary of Fig.~\ref{fig:fig1}. \\
 \\
Moreover, the dynamical time scale is the (relatively)
slowest at the turning point, of order several $fm/c$ at these energies, such that
equilibrium might be approached and the evolution would exhibit a sensitivity to
thermodynamic features such as a critical point, and to bulk matter phenomena like a first
order phase transition. Please note that the $A+A$ dynamical evolution never proceeds on a
''line'' in any phase diagram, nor does it collectively reach any ''point'' in the ($T,
\mu_B$) plane, due to the relatively unfavourable surface to volume ratio creating density
gradients in the ''fireball'' system - the evolution proceeds within a relatively broad
band of local conditions. One would thus not expect a sharply singular behaviour of any
observable with $\sqrt{s}$.

\begin{figure}
\begin{center}
\includegraphics[width=8cm]{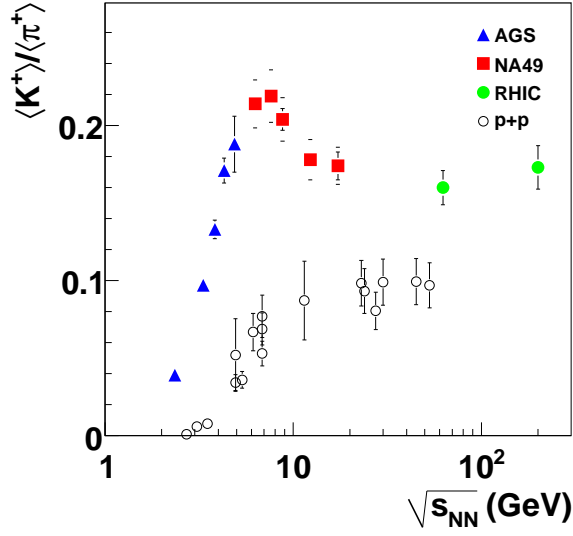}
\caption{Energy dependence of the $K^+/\pi^+$ ratio in central $A+A$ and in $p+p$
collisions, from AGS to RHIC energies \cite{22,30}.}
\label{fig:fig5}
\end{center}
\end{figure}

\begin{figure}
\begin{center}
\includegraphics[width=5.8cm]{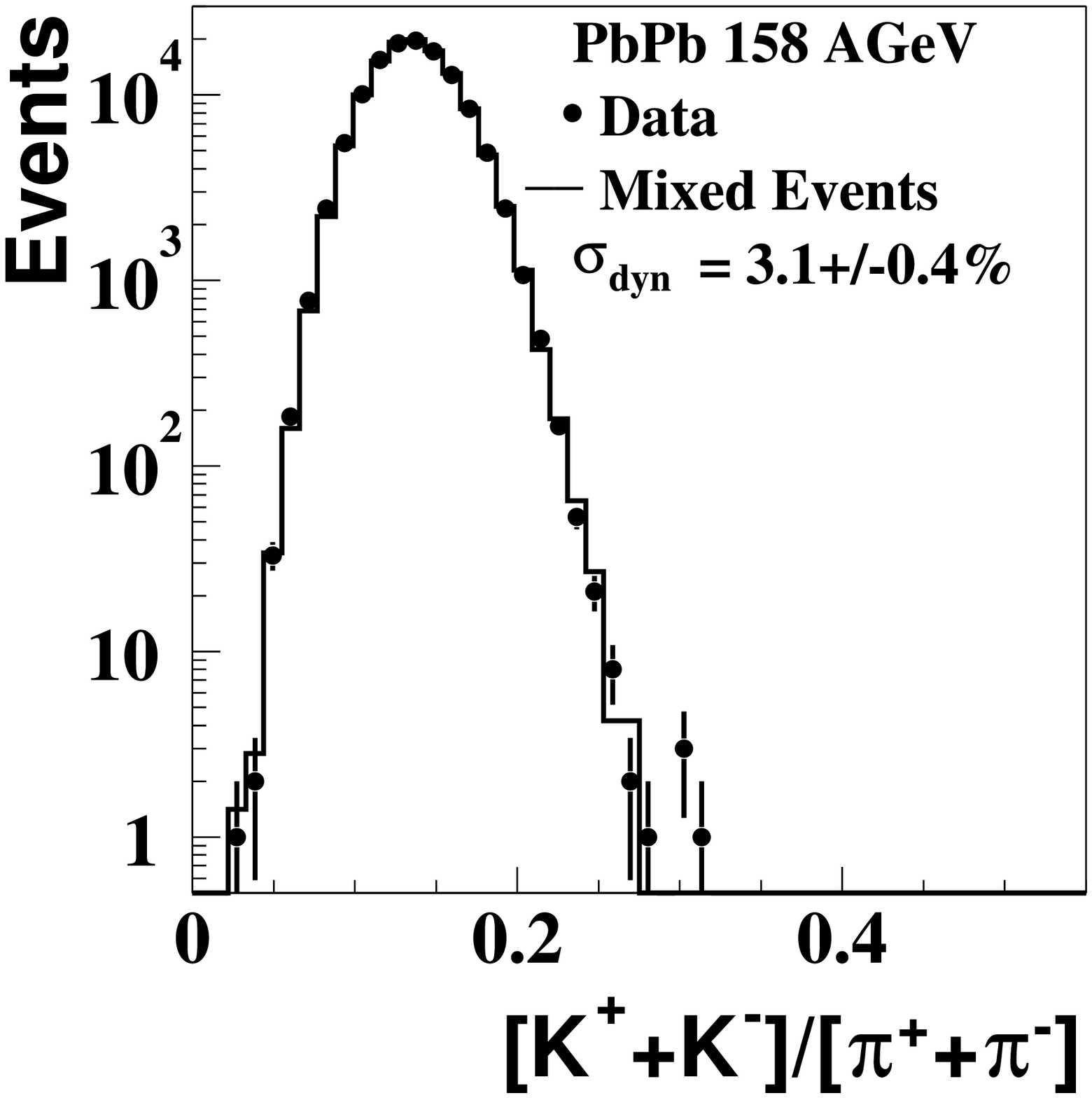}
\includegraphics[width=7cm]{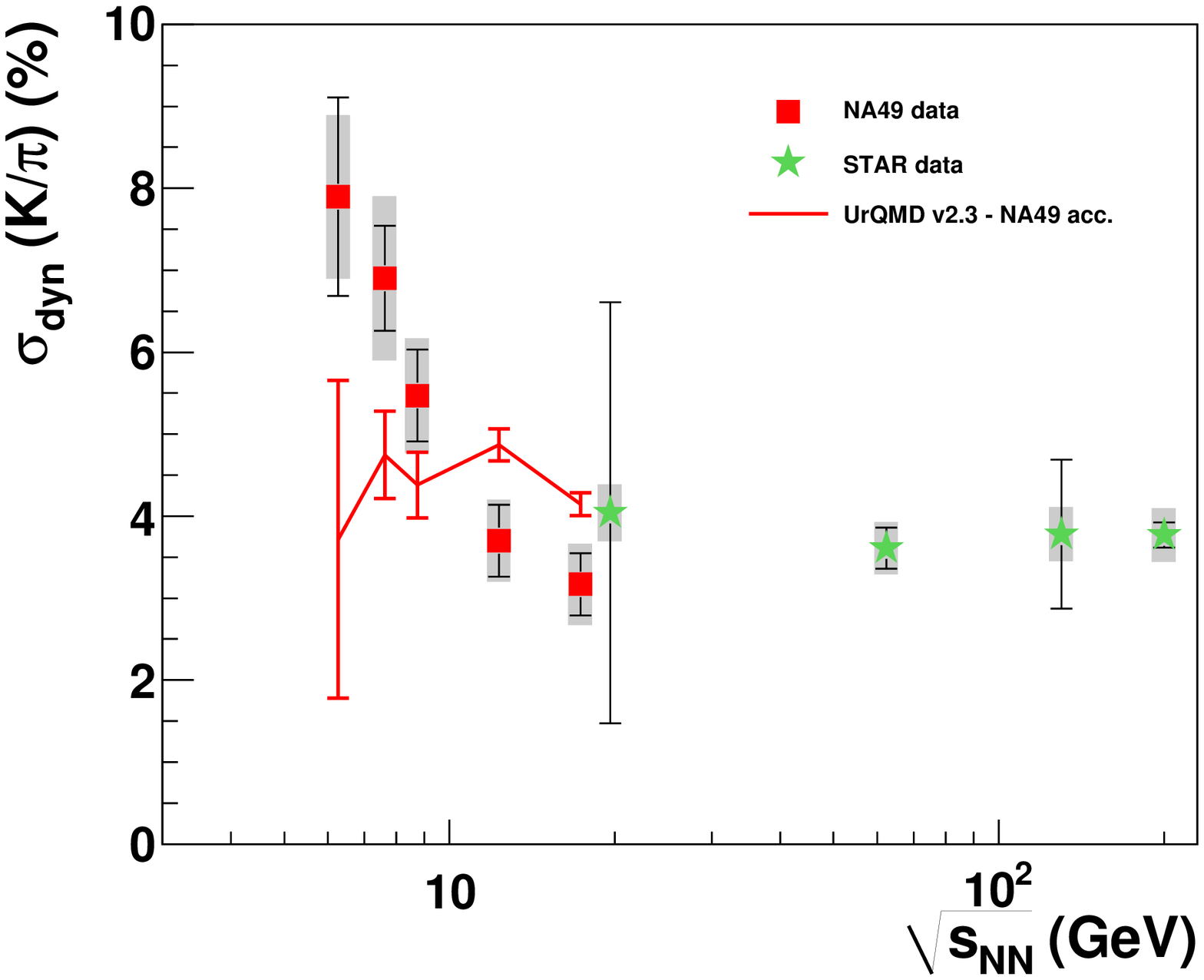}
\caption{Left: Event by event fluctuation of the charged kaon to charged pion multiplicity
in central collisions of $Pb+Pb$ at $\sqrt{s}=17.3 \: GeV$, compared with background from
mixed events (histogram) \cite{33}. \newline Right: energy dependence of $\sigma_{dyn}$
for kaon to pion fluctuation at SPS \cite{35} and RHIC \cite{34} energies, compared to
UrQMD prediction.}
\label{fig:fig6}
\end{center}
\end{figure}

\noindent Seen in this light, the relatively sharp peak observed by NA49 in the
$K^+/\pi^+$ excitation function \cite{30} is remarkable. Fig.~\ref{fig:fig5} shows the $\sqrt{s}$
dependence of this ratio, from threshold behaviour (at AGS energy) to peak at low SPS
energy, with decline toward top SPS and RHIC energy. This is probably the sharpest
dependence on $\sqrt{s}$ that could be expected. Is it a signature of ''onset of
deconfinement''? We will turn to this open problem below, and in the next section, but
note here that Fig.~\ref{fig:fig5} also illustrates the phenomenon called ''strangeness enhancement in $A+A$''. Note that $K^+$ represents the bulk of (anti-)strangeness production, and that the $K^+/\pi^+$ ratio thus approaches the global strangeness to entropy ratio. Fig.~\ref{fig:fig5} also illustrates the data for elementary minimum bias $p+p$ collisions which exhibit no peak structure. The dramatic overall increase in central $A+A$ collisions has been understood as a consequence of the formation of a large, quantum number coherent interaction volume in $A+A$ collisions, absent in $p+p$ interactions \cite{31}. Within the terminology of the statistical hadronization model, this effect is reflected in a transition from a canonical to a grand canonical Gibbs ensemble description \cite{31}, where ''canonical strangeness suppression'' in elementary, small volume collisions is seen to arise from strict quantum number conservation constraints at the microscopic level, fading away in large, coherent volume solutions, corresponding to the $A+A$ case, which lead to strangeness saturation. The $K^+/\pi^+$ signal in Fig.~\ref{fig:fig5} thus tells us that $A+A$ collisions create a large quantum number coherent ''fireball'' volume {\bf before hadronic freeze-out sets in}
\cite{32}. \\
 \\
The $K/\pi$ ratio should also be ''robust'' insofar as it will survive final state interaction during the late stages of hadronic expansion, during which strangeness and entropy are approximately conserved. This motivates the analysis of its fluctuation at the event-by-event level \cite{33,34,35}. Within finite sub-intervals of rapidity space this might be caused by the lattice quark density fluctuations shown in Fig.~\ref{fig:fig3}. As shown in Fig.~\ref{fig:fig6} (left) the initial measurement \cite{33} of the quantity $\sigma_{dyn}(\frac{K}{\pi})$ gave a very small result, the signal being almost fully exhausted by the mixed background. Fig.~\ref{fig:fig6} (right) shows that this remains so at RHIC energy \cite{34} but that $\sigma_{dyn}$ exhibits a significant rise toward low SPS energy \cite{35}, where $\mu_B \rightarrow 450 \: MeV$, the region of interest from Figs.~\ref{fig:fig2} and~\ref{fig:fig3}. A simulation with the UrQMD transport model shows no such dependence. Randrup and collaborators have suggested \cite{36} to approach, both the ensemble average $K/\pi \:
(\equiv K^+ +K^-/\pi^+ + \pi^-)$ ratio, and its event-wise fluctuation, not as a critical
point consequence but as effects of the spinodal decomposition mechanism that could
develop during a first order phase transition. The global grand canonical fireball
(constrained by charge and strangeness conservation on average) would decompose into
separately expanding ''bubbles'' in phase space which could carry nonzero net strangeness.
Fig.~\ref{fig:fig7} shows that the ensemble average kaon yields (left) exhibit no difference from the grand canonical global values (as could have been expected from the percolation models for strangeness one \cite{32}) but that the event by event $K^+$ multiplicity variance
does indeed grow well beyond the GC value, toward high $\mu_B$, similar to the data in
Fig.~\ref{fig:fig6}.

\noindent We shall return to these observables in the next sections but end, at this
point, our introductory tour de horizon of the field - as is stood at the previous Firenze
CPOD workshop \cite{37}. We now turn to the present physics of the QCD phase diagram.\\

\section{Recent Developments}

\subsection{The critical Point Focusing Effect}

\noindent Asakawa and Nonaka have developed a hydrodynamical model which implements a
critical region \cite{38,39}, causing a sensational attractor or focusing effect on the
dynamical trajectories of the system expansion at different $\sqrt{s}$. Their work is
illustrated in Figs.~\ref{fig:fig8} and~\ref{fig:fig9}. Fig.~\ref{fig:fig8} (right) gives dynamical trajectories in the absence of a CEP but including a first order phase transition taking place at the parton-hadron boundary line. \\
\begin{figure}
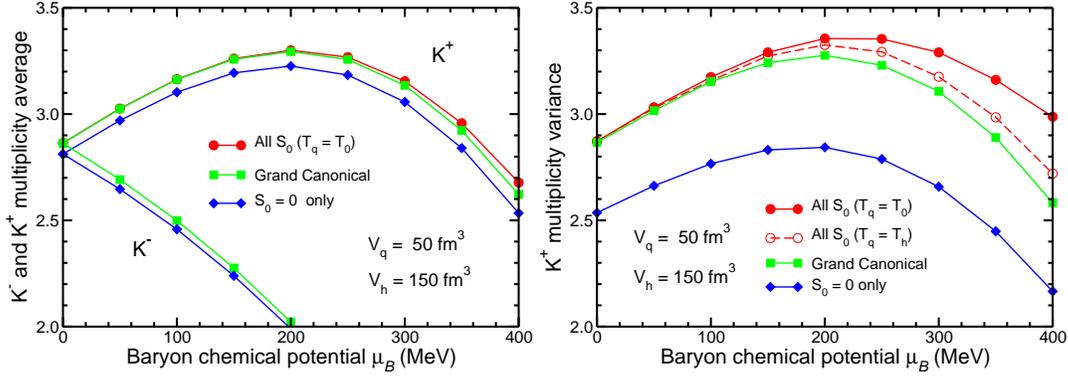

\begin{center}
\includegraphics[width=7cm]{07A_fig-2_0509030.eps}
\includegraphics[width=7cm]{07B_fig-3_0509030.eps}
\caption{Effect of hydrochemical spinodal decomposition on kaon production in central
$A+A$ collisions \cite{36}. Comparison to a single grand canonical fireball is made, with
increasing $\mu_B$, for the average multiplicity (left) and for the $K^+$ multiplicity
fluctuation width.}
\label{fig:fig7}
\end{center}
\end{figure} \\
It is important to recall that in reality an $A+A$ central collision
creates, not such lines but relatively broad bands in the $[T, \: \mu_B]$ plane but we
infer, nevertheless, that the regions in which the equilibrium period of the fireball
expansion enters the figure (from above) are monotonously shifting (upward and) to the
left with increasing $\sqrt{s}$. The influence of the confinement mechanism considered
leads to back-bending and some further reshuffling (characteristic of the phase
transformation type implied) of trajectories. There is no hadronic freeze-out considered
in this model. But we know from Fig.~\ref{fig:fig1} that the data are arranged in the close vicinity of
the coexistence line, falling off at large $\mu_B$. It is important, now, to note that
each expansion trajectory has to pass through such a chemical freeze-out ''point''. So,
from this version of the hydro model we infer a smooth, monotonous relationship between
$\sqrt{s}$ and freeze-out points, and this is, in fact, the result of the systematic
analysis with the statistical model \cite{16,17,18,40,41}. But we {\bf also} learn that
this relationship must be sensitive to the position of the coexistence line, and to the
nature of the phase transformation.

\noindent This is true, in particular, if a CEP focusing mechanism is introduced, as
illustrated in Fig.~\ref{fig:fig8} (left). The trajectories enter, at model-time zero, at $T=200 \: MeV$ and $\mu_B=145, \: 461$ and $560 \: MeV$, respectively, similar to the case with no CEP, but get substantially defocused. The relationship between $\sqrt{s}$ and freeze-out
$\mu_B$ would be dramatically affected. This is further illustrated  \cite{39} in Fig.~\ref{fig:fig9}, which shows the expansion trajectory in the close vicinity of an entropy to baryon number density ratio (conserved during hydro-expansion, thus a ''label'' of the individual trajectory) of about 25. The {\bf same} chemical freeze-out point is reached by 3
different trajectories corresponding to different $\sqrt{s}$ and assumed mechanism for
crossing of the phase boundary: a first order transition without (FO) and with (QCP)
presence of a CEP, and a crossover transition.

\noindent Without turning to discuss the degree of realism of this model (please consult
\cite{38,39}) we conclude that the freeze-out points in Fig.~\ref{fig:fig1} need to be revisited. \\
\begin{figure}
\begin{center}
\includegraphics[width=7cm]{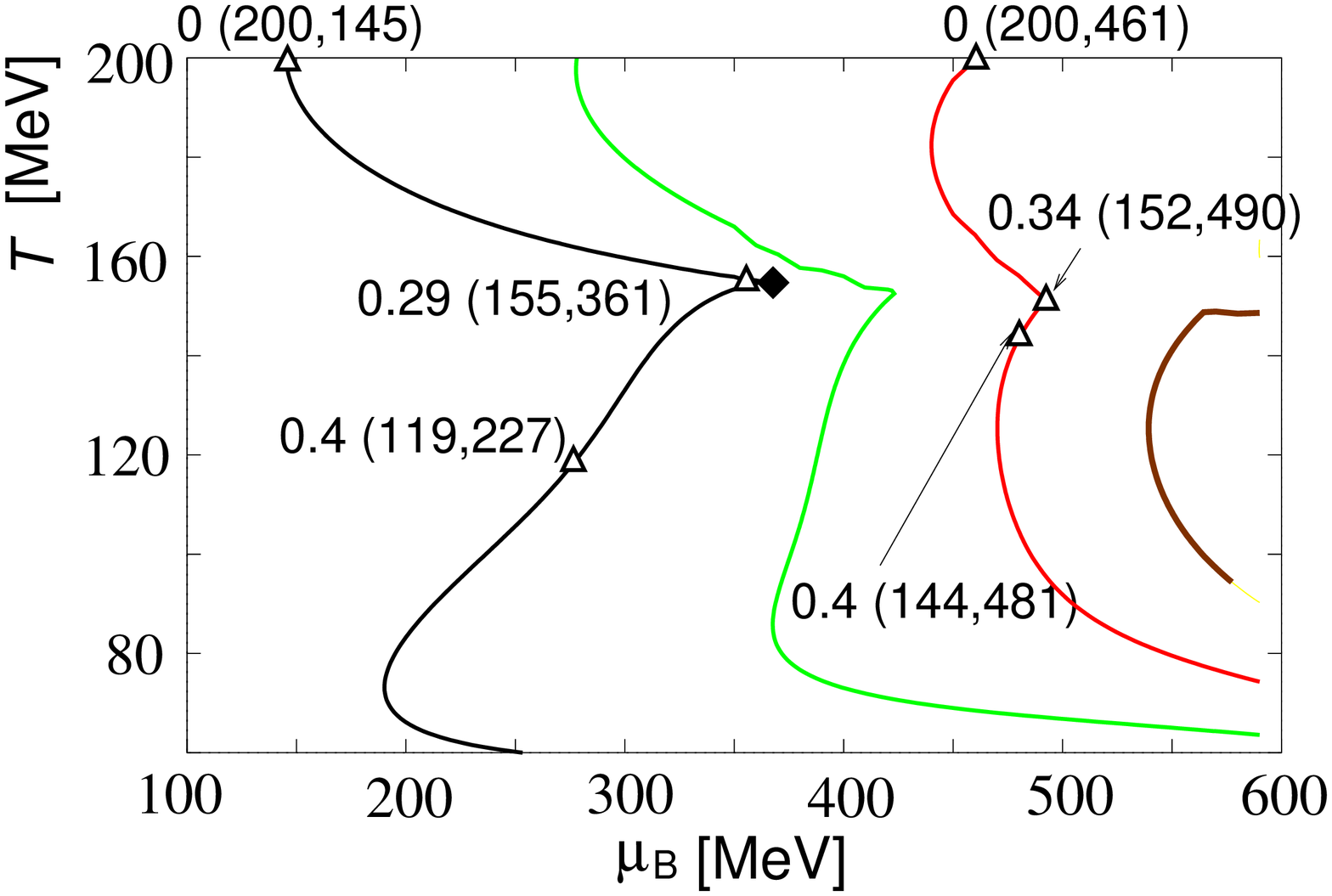}
\includegraphics[width=7cm]{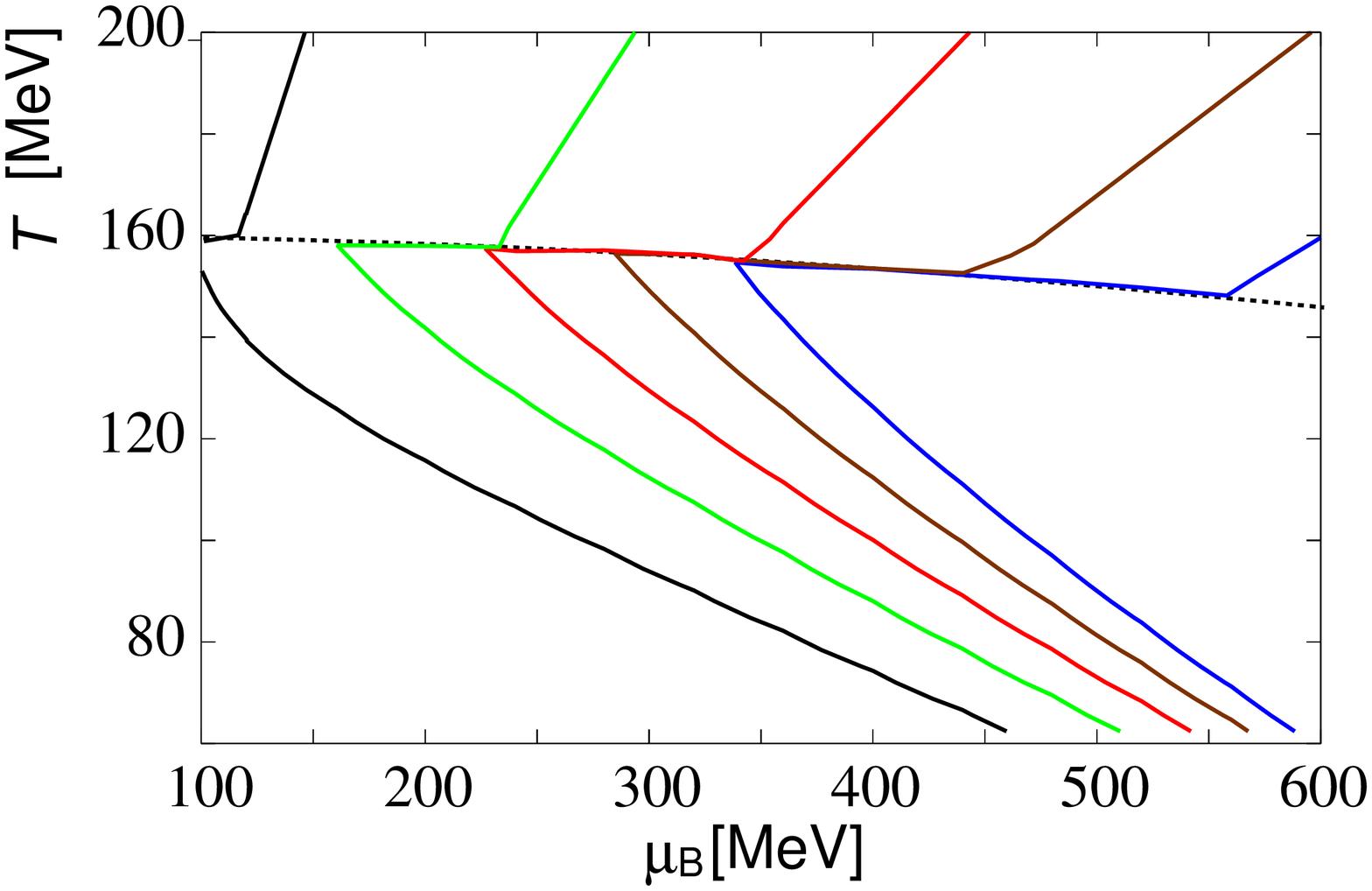}
\caption{Left: effect of a critical point on the hydrodynamic expansion trajectories in
central $A+A$ collisions at various energies \cite{38}. \newline Right: the same initial
trajectories without a critical point but showing the effect of a first order phase
transition in the EOS.}
\label{fig:fig8}
\end{center}
\end{figure}
\subsection{Hadrochemical Freeze-out at High {\bf $\mu_B$}}

Turning to the hadrochemical freeze-out points in Fig.~\ref{fig:fig1}, we note, first of all, that the statistical Grand Canonical ensemble of hadrons and resonances can not, in its ideality,
be referred to the realistic situation encountered in central $A+A$ collisions without
substantial additional assumptions and approximations \cite{21,22}. Think e.g. of rapidity
windows, surface ''corona'' effects, canonical strangeness suppression, finite hadron
size, etc.. Thus, each of the various ''schools'' of statistical model analysis has, over
time, changed (improved) its technical details, and, moreover, different choices of
dealing with such second order concerns have been persued. Some scatter of the published
[$T, \: \mu$] freeze-out points for each $\sqrt{s}$ is the consequence and, taken
together, they convey the impression of a band rather than a curve; this band has then
been well described by a smooth parametrization \cite{23,40,41}, the ''freeze-out curve''.

\noindent However, toward our present concern about non-smooth excursions from the
freeze-out curve, we take the opposite approach. We illustrate in Fig.~\ref{fig:fig10} {\bf
exclusively} the freeze-out points from a single most recent systematical investigation of
all available data \cite{23}. The overall trend remains but excursions are obvious,
perhaps significant in view e.g. of the CEP trajectory re-focusing discussed above. These
results are the consequence of an attempt to describe the $K^+/\pi^+$ peak of Fig.~\ref{fig:fig6} that we illustrate in Fig.~\ref{fig:fig11}. It is the novel ingredient of analysis \cite{23}, effective here, to consider more explicitly the energy dependence of the hadronic resonance spectrum that enters the Grand Canonical hadron plus resonance Gibbs ensemble. The model now captures the decline in the $K^+/\pi^+$ and $\Lambda/\pi^-$ signals at $\sqrt{s}$ above about $8 \: GeV$, the sharp peak structure in $K^+/\pi^+$ still elusive. However, it is also obvious from Fig.~\ref{fig:fig11} that the data in the SPS energy range need significant improvement. A far higher accuracy is also of crucial importance to our next topic.

\noindent We return to the hadronic freeze-out points, smooth or not in Fig.~\ref{fig:fig1} and~\ref{fig:fig10}, and focus at their fall-off from the confinement line suggested/conjectured by lattice QCD. Note that the term freeze-''out'' refers to the Gibbs ensemble of on-shell hadrons and resonances in vacuum, as it is employed in the statistical model of hadronization. There is no mean field, and zero virtuality. The hadron gas is still interacting due to the presence of the full resonance spectrum, but the number distribution of species (after resonance decay) stays constant (frozen-in) from this point on. Note that $T=160-170\:
MeV$ represents a relatively cold medium, with small average inelasticity. This is
illustrated in Fig.~\ref{fig:fig12} (top) by a hydro plus UrQMD calculation by Bass and Dumitru \cite{42} for central $Au+Au$ collisions at top RHIC energy. 
\begin{figure}
\begin{center}
\includegraphics[width=9cm]{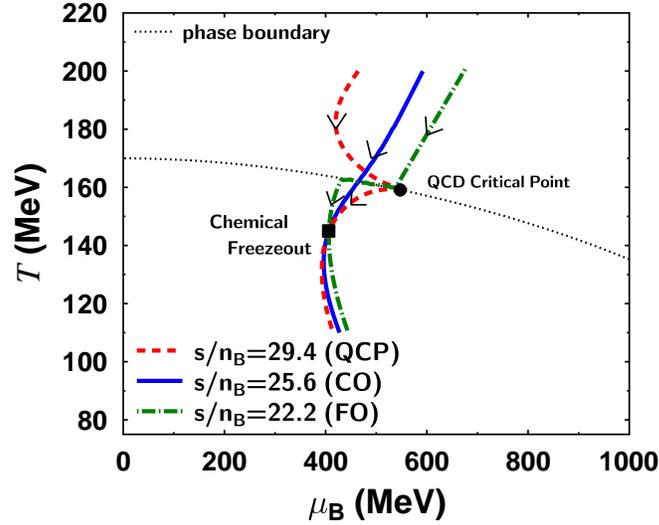}
\caption{Same model as Fig.~8 \cite{39}: three initial expansion trajectories
characterized by their entropy to baryon density ratio $S/n_B$, experiencing the effect of
a critical point, a first order, and a cross-over phase transition, respectively, end up
in a common freeze-out point.}
\label{fig:fig9}
\end{center}
\end{figure}
The yield distribution after Cooper-Frye termination of the hydrodynamical evolution at $T=170 \: MeV$ is compared with the outcome if a UrQMD hadron-resonance cascade expansion phase is attached. It causes only minor changes of the population, which thus stays frozen-in, for the case of
hadronization of $T=155 \: MeV$ and $\mu_B \approx 20 \: MeV$. A similar observation is
made at SPS energy \cite{43}. Fig.~\ref{fig:fig12} (bottom) illustrates the same procedure at
$\sqrt{s}=8.7 \: GeV$ for the case of hadronization at $150 \: MeV$ and $\mu_B=350 \:
MeV$. These calculations illustrate the well-known fact that the chemical relaxation time
of a binary hadron-resonance cascade mechanism is far too long \cite{44} to achieve
equilibrium or, turning around, to abolish an equilibrium previously installed by the
hadronization process \cite{20,21,22,45}.

\noindent We are led to conclude on an apparent impasse: if confinement and hadronization
occurs at the phase boundary, also at higher $\mu_B$, there is no simple transport
mechanism that would carry the chemical equilibrium from hadronization at, say, $150-160
\: MeV$ downward to the {\bf observed} hadronic freeze-out equilibrium at, say, $135-145
\: MeV$.

\noindent One way out: the data entering the statistical model analysis are insufficient
at $\sqrt{s}\le 9 \: GeV, \: \mu_B \ge 300 \: MeV$, because the phi-meson and higher
hyperon multiplicities are inaccurate, or missing. In other words, chemical equilibrium
might in fact be incomplete or missing, the freeze-out points at high $\mu_B$ thus
appearing illegitimately  on the equilibrium phase diagram of Fig.~\ref{fig:fig1}. However, there are no indications of such a deviation from equilibrium as far down as $\sqrt{s}= 8.7 \: GeV,\: \mu_B=380 \: MeV$, as is shown in Fig.~\ref{fig:fig13} by the fit \cite{41} to the corresponding SPS, $Pb+Pb$ central collisions data of NA49. Lower energies will hopefully be accessible during the planned RHIC runs at low energy. \\
\begin{figure}
\begin{center}
\includegraphics[width=9cm]{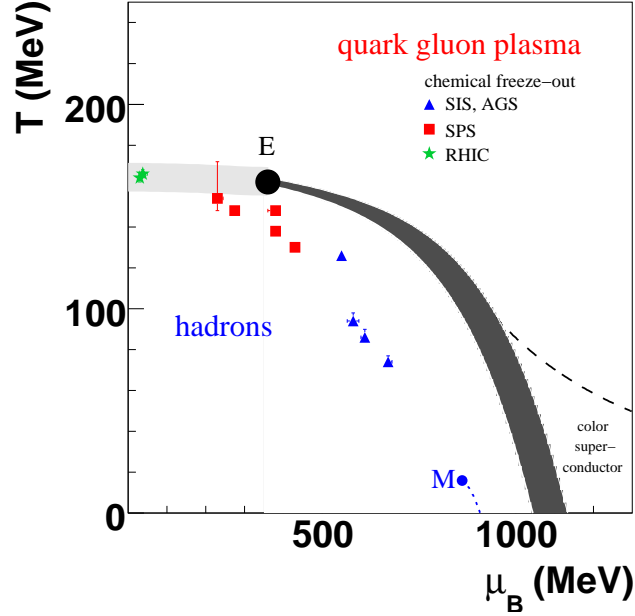}
\caption{The QCD phase diagram of Fig.~1 but with statistical model hadron freeze-out points from an analysis \cite{23} including an energy dependence of the hadronicresonance spectrum.}
\label{fig:fig10}
\end{center}
\end{figure}\\
The alternative: there should be a new {\bf state} of QCD intervening between
the color confinement and the hadronic freeze-out curves. This conclusion is also hinted
at by the observation \cite{23} that the energy density along the lattice-inspired color
confinement line far exceeds the canonical $1 \: GeV/fm^3$, at $\mu_B \rightarrow 500\:
MeV$. Not a likely hadronization condition. So that it is tempting to identify the
freeze-out points as the Hagedorn limiting temperature \cite{46,47}, i.e. as the limit of
existence of hadron-resonance matter, at high $\mu_B$. Finally, there does actually even
exist a candidate of an intervening, new QCD phase, conjectured by McLerran, Pisarski and
collaborators \cite{24}, and named ''quarkyonic matter''. Fig.~\ref{fig:fig14} gives a {\it sketch} of the corresponding phase diagram of QCD, which features a phase boundary of color
confinement and a further one of hadronization from quarkyonium to hadrons plus resonances
(about which nothing is known in detail as of yet). This would finally explain the origin
of the chemical equilibrium, implied by the hadronic freeze-out points at high $\mu_B$: to
closely coincide with the latter phase boundary \cite{24}. In keeping with the observation
\cite{20,21,22,45} that hadrons do not dynamically acquire chemical equilibrium but are
''born into it'' \cite{19,20} by a phase transition which enforces phase space dominance
via the emerging hadron-resonance spectrum \cite{48}. In this view, the apparent
equilibrium in $A+A$ freeze-out requires synchronization by a global change of degrees of
freedom \cite{48}, unlike freeze-out from a smooth dilution process caused by expansion,
such as the cosmological synthesis of light nuclei, or the final momentum space decoupling
of diluting hadronic cascades, which necessarily lead to {\bf sequential} freeze-out \cite{22}, in order of total cross sections.

\noindent It will be important to measure complete hadron species multiplicity
distributions, down to the anti-Omega, at energies as low as $\sqrt{s}=6 \: GeV$ (where
this multiplicity is as low as 0.02). A challenging experiment task, given the low
luminosities expected here, at RHIC, but the only test of the quarkyonium hypothesis known thus far, as it addresses the domain of $\mu_B \rightarrow 500 \: MeV$ where the present, preliminary freeze-out points fall really far below the lattice QCD confinement line
estimate (Fig.~\ref{fig:fig1}).\\
\begin{figure}
\begin{center}
\includegraphics[width=7cm]{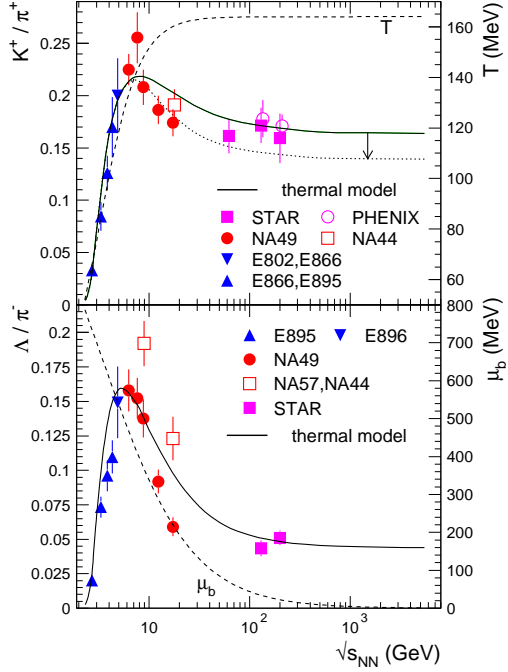}
\caption{Energy dependence of the $K^+/\pi^+$ and $\Lambda/\pi^+$ ratio with statistical
model description \cite{23} as in Fig.~10.}
\label{fig:fig11}
\end{center}
\end{figure}

\section{Fluctuation Signals from Event-by-Event Physics Analysis}
Central collisions of heavy nuclei create bulk hadrons, at $p_T<2 \: GeV$, with a
multiplicity per event that is high enough to perform analysis of physics observables at
the level of single events. Experiments such as NA49 at the SPS, and STAR, PHOBOS at RHIC were designed, specifically, for this task. Most copiously produced charged hadrons are $K^+$ and $K^-$, protons (and antiprotons at RHIC), and charged pions, with recorded average multiplicities per event ranging from about 5 for $K^+$ at low SPS energy, to several hundreds of pions at the top RHIC energy. Each central $Pb+Pb$ or $Au+Au$ collision thus represents, in statistical model terminology, a ''grand micro-canonical ensemble'', where the term ''grand'' indicates the large production volume, but the term ''micro'' reminds us of the fact that each individual large volume event must overall strictly conserve electric, baryonic and strangeness charge conservation. This complicates the seemingly straight-forward idea of event by event analysis, as far as conserved quantities are concerned \cite{49}. The first considerations of fluctuations, occurring as a consequence of a QCD critical point \cite{26}, referred to non-conserved quantities: eventwise analysis of multiplicity density, and transverse momentum fluctuations. We shall turn to the latter observable below, then to continue with hadron ratio and higher moment fluctuations, such as kurtosis, which make contact to the recent lattice QCD calculations. \\
\begin{figure}
\begin{center}
\includegraphics[width=7cm]{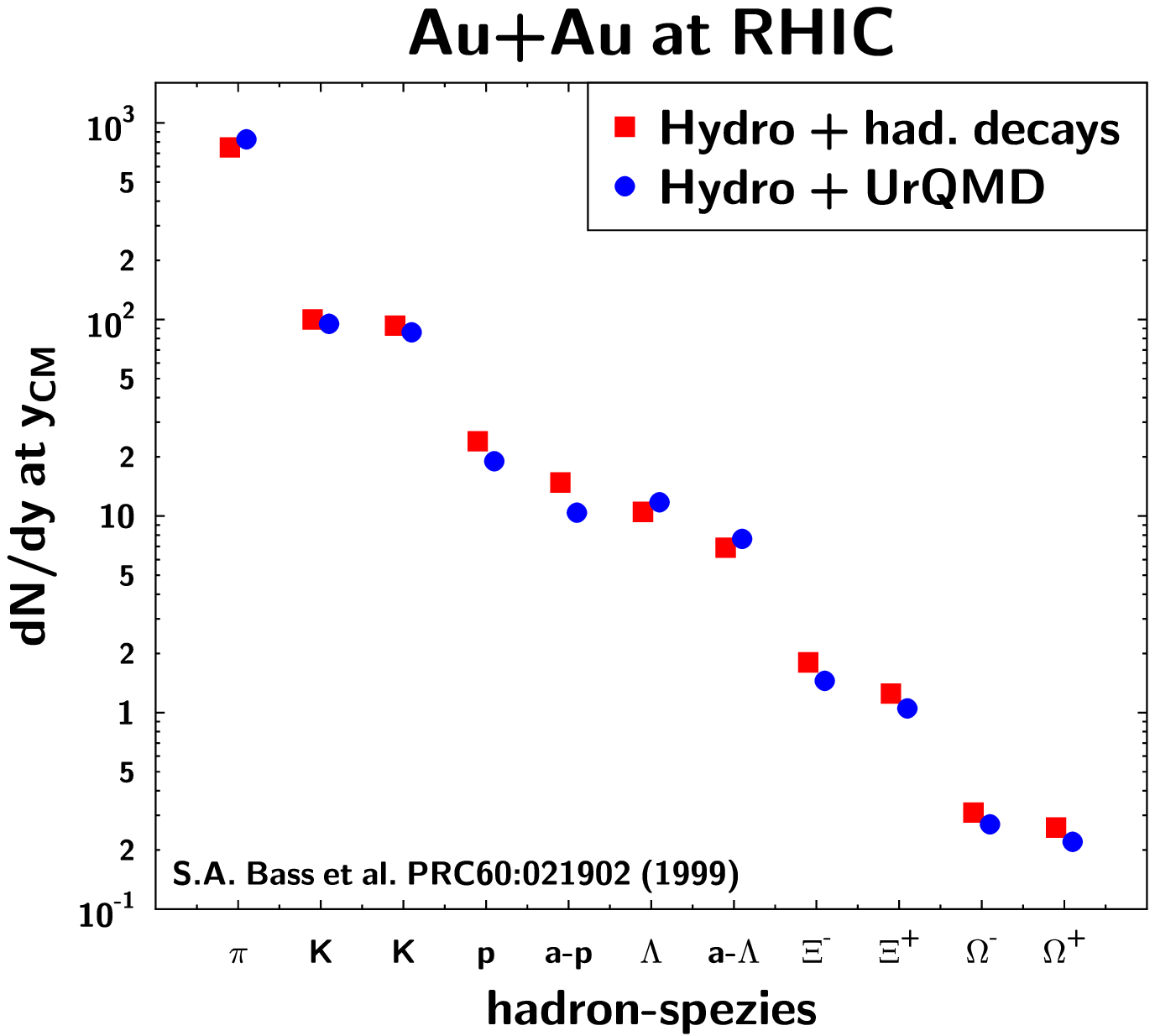} \\
\includegraphics[width=7cm]{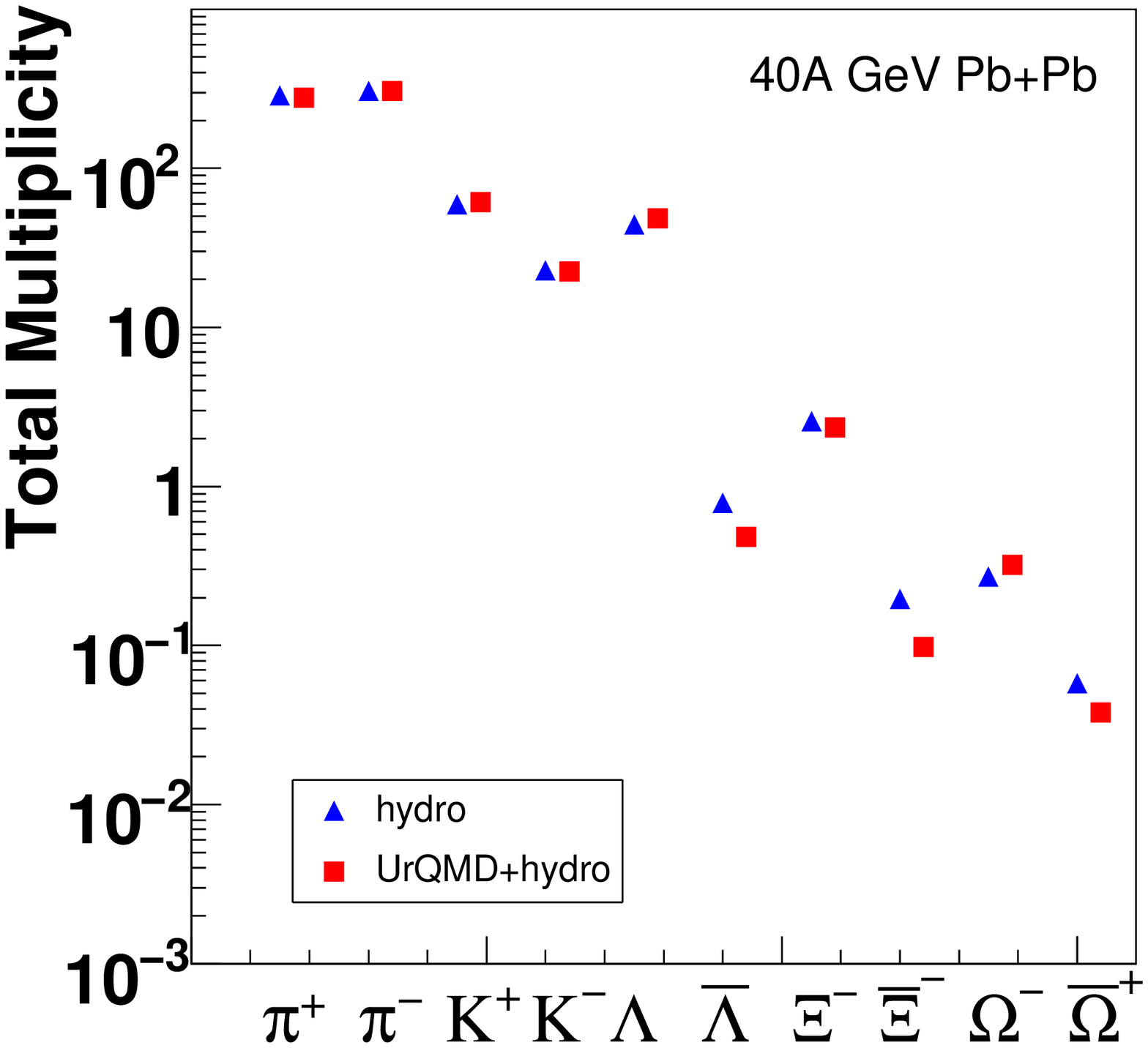}
\caption{Top: effect of the hadronic expansion period on the hadro-chemical composition
established at hadronization of a hydrodynamic partonic expansion, for central $Au+Au$ at top RHIC energy \cite{42}. \newline Botton: same comparison \cite{43} for $Pb+Pb$ at SPS energy 40 $AGeV$ ($\sqrt{s}=8.7 \: GeV$).}
\label{fig:fig12}
\end{center}
\end{figure}

\subsection{Transverse Momentum Fluctuations}
The classical, pioneering article of Stephanov, Shuryak and Rajagopal \cite{26} considered
a critical point of QCD within the framework of the chiral symmetry restoration transition
of QCD. Complementary to the deconfinement transition, between bound color singlets and
finite color mobility, this transition addresses the transition between massive hadrons
(with mass caused by non-perturbative parton condensates in the non-perturbative vacuum)
and approximately massless light quarks. At low $\mu_B$ these two QCD transitions appear
to occur synchronously \cite{14,15}. The order parameter of the chiral phase transition is
the mass of the sigma field meson which approaches zero at the critical point, causing
''critical'' field fluctuation \cite{26}. The linkage to experimentally observable
quantities consists in the $\sigma \rightarrow 2 \pi$ decay near threshold, causing
charged particle multiplicity density fluctuations \cite{50} as well as transverse
momentum pair correlation, and overall transverse momentum fluctuation, in the vicinity of
a QCD critical point. \\
\begin{figure}
\begin{center}
\includegraphics[width=9cm]{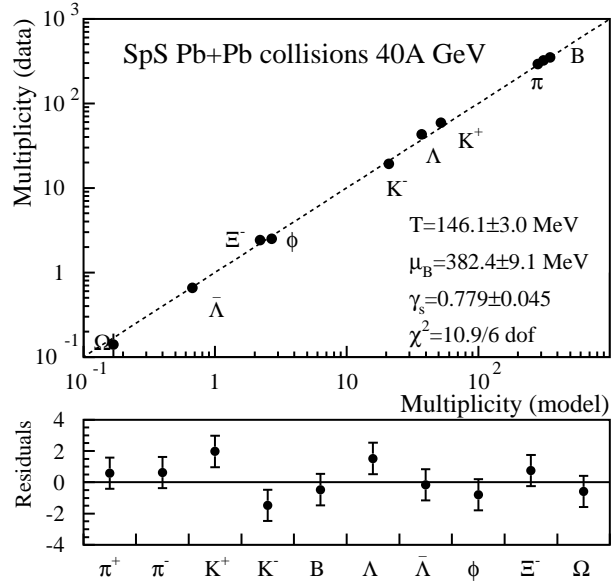}
\caption{Statistical model fit \cite{41} to hadronic multiplicities obtained in central
$Pb+Pb$ collisions at $40 \: GeV \: (\sqrt{s}=8.7 \: GeV$) by NA49.}
\label{fig:fig13}
\end{center}
\end{figure}

\noindent These predictions have inspired numerous experimental investigations, first
begun at SPS energy. Fig.~\ref{fig:fig15} shows \cite{51} the distribution of the eventwise average charged particle $p_T$ in central collisions of $Pb+Pb$ at top SPS energy, $\sqrt{s}=17.3 \: GeV$. A perfect Gaussian. The histogram represents the signal derived from randomized Monte Carlo sampling of tracks throughout the recorded event ensemble: the ''mixed background''. The real event data show no significant deviation from the mixed background, over four orders of magnitude: a first ''null''-result, at this energy.

\noindent We shall see in the next section that such signal comparisons to a mixed
background can also lead to significant differences, for other observables. But we note,
for now, that this mode of analysis may be intuitively convincing but leads to
considerable difficulty, as far as quantitative comparison to theoretical models is
concerned. The mixed background distribution reflects non-trivially convolution of
experimental biases stemming from acceptance, tracking and particle identification
efficiencies, momentum resolution limitations etc., which also might affect the single
event observation in a way different from the ensemble averaged background, due to impact
parameter fluctuation, and to event-by-event efficiency variation, correlated e.g. to the
eventwise multiplicity density distribution. Thus, several alternative measures of
fluctuation have been suggested. One can base a signal of dynamical $p_T$ fluctuation on
the binary {\it correlation} of particle transverse momenta in a given event i.e. on the
covariance ($p_{Ti} \: p_{Tj}$) of particles $i, \: j$ in one event \cite{52}. Of course,
the covariance receives contributions from sources beyond our present concern, i.e.
Bose-Einstein correlation, flow and jets (the jet activity becomes prominent at high
$\sqrt{s}$, and will dominate the $p_T$ fluctuation signal at the LHC). In covariance
analysis, the dynamical $p_T$ fluctuation (of whatever origin) is recovered via its effect
on correlations among the transverse momentum of particles.  \\
\begin{figure}
\begin{center}
\includegraphics[width=9cm]{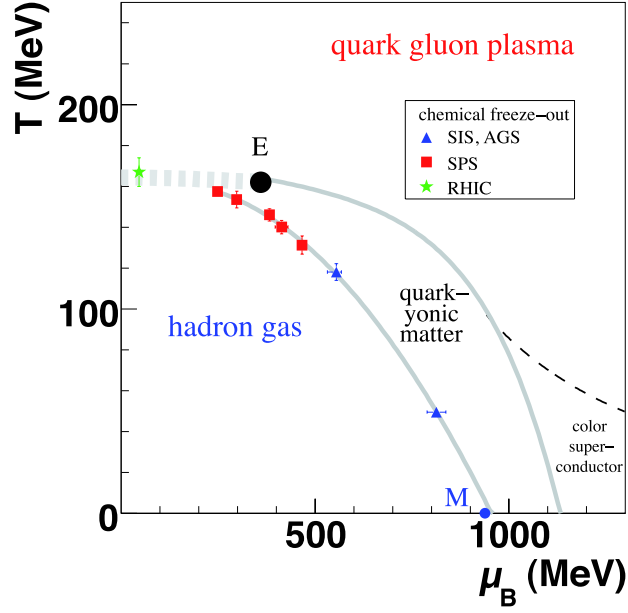}
\caption{Sketch of the QCD phase diagram including the hypothetical quarkyonium phase
\cite{24}. The hadronic freeze-out points now follow the hadronization line.}
\label{fig:fig14}
\end{center}
\end{figure} \\
Such correlations can be quantified employing the two-particle $p_T$ correlator \cite{52,53}
\begin{equation}
(\Delta p_{Ti} \Delta p_{Tj})=\frac{1}{M_{pairs}} \sum^n _{k=1} \: \sum^{N(k)} _{i=1} \:
\sum ^{N(k)} _{j=i+1} \Delta p_{Ti} \Delta p_{Tj}
\end{equation}
where $M_{pairs}$ is the total number of track pairs of the events $k$ contained in the
entire ensemble of $n$ events, $N(k)$ is the number of tracks in event $k$, and $\Delta
p_{Ti}=p_{Ti}-\overline{p}_T$ where $\overline{p}_T$ is the global ensemble mean $p_T$.
The normalized dynamical fluctuation is then expressed \cite{52} as
\begin{equation}
\sigma(p_T)_{dyn} = \sqrt{(\Delta p_{Ti} \Delta p_{Tj})} / \overline{p}_T
\end{equation}
It is zero for uncorrelated particle emission.

\noindent Fig.~\ref{fig:fig16} shows the analysis of $p_T$ fluctuations based on the $p_T$ correlator, for central $Pb+Au$ SPS collisions by CERES \cite{53} and for central $Au+Au$ at four RHIC energies by STAR \cite{52}. The signal is at the 1\% level at all $\sqrt{s}$, with no hint at critical point phenomena. Its small but finite size could arise from a multitude of
sources, e.g. Bose-Einstein statistics, Coulomb or flow effects, mini-jet-formation, but
also from experimental conditions such as two-track resolution limits \cite{53}. NA49 has
employed \cite{54} the measure $\Phi(p_T)$ defined as \cite{55}
\begin{equation}
\Phi(p_T)=\sqrt{\frac{<Z^2>}{<N>}}- \sqrt{\overline{z^2}}
\end{equation}

\noindent where $z_i=p_{Ti}-\overline{p}_T$, with $\overline {p}_T$ the overall inclusive
average; for each event one calculates $Z=\sum_n z_i$, with $N$ the multiplicity of the
event. With the second term the trivial independent particle emission fluctuation is
subtracted out, i.e. $\Phi$ vanishes if this is all. \\
\begin{figure}
\begin{center}
\includegraphics[width=7cm]{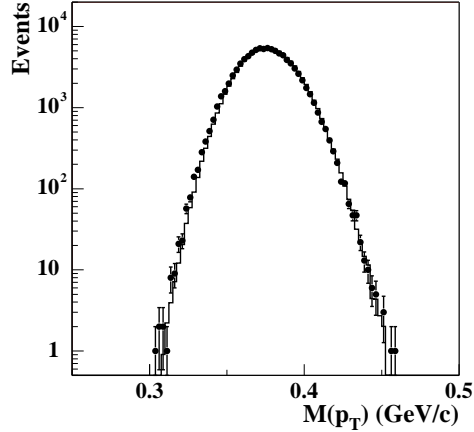}
\caption{Distribution of event average $p_T$ in central $Pb+Pb$ collisions at
$\sqrt{s}=17.3 \: GeV$, compared to mixed background histogram \cite{51}.}
\label{fig:fig15}
\end{center}
\end{figure} \\
Fig.~\ref{fig:fig17} shows the NA49 result \cite{54} for central $Pb+Pb$ collisions, and indeed $\Phi$ is compatible with zero throughout. The SPS energies, from $158 \: AGeV$ to $20 \: AGeV$, are represented here by the prevailing baryochemical potential values \cite{41}. The figure includes a study of the effects that might be expected from a critical point, the position and width $\xi$ of which are taken from the lattice calculations \cite{56,57}. The expected signal amplitude is then calculated \cite{58} in the framework of ref. \cite{26}. It is relatively small for realistic $\xi$ but the data do not allow for a final conclusion; at least there is clearly no extremely long range of correlation, such as $6 \: fm$. We see that a
substantial effort is required to conclude on this challenging critical point search
proposal. Furthermore, pion momentum fluctuations imprinted at the phase boundary might be dissipated away, and thus ''thermalized''to the final momentum space freeze-out, at about $T=100 \: MeV$, as a pion experiences about 5 rescatterings on average during the final hadronic cascade \cite{42}. On the other hand a hadro-chemical $K/\pi$ or $K/p$ ratio fluctuation would be preserved throughout the cascade as can be inferred from Fig.~\ref{fig:fig11}.

\subsection{$K/\pi$ Fluctuations}

We showed in Fig.~\ref{fig:fig6} the first results concerning $K/\pi$ (strangeness to entropy)
fluctuations. We are looking for a reflection of the light and strange quark
susceptibility maxima at $\mu_B=3 \:T_c$ shown in Fig.~\ref{fig:fig3}, or of spinodal decomposition (Fig.~\ref{fig:fig7}). The fluctuations are small from top SPS to RHIC energy in central collisions of $Pb+Pb$ or $Au+Au$, but exhibit a remarkable rise toward the lowest SPS energies which can not be reproduced by UrQMD calculations \cite{35}. The employed observable, $\sigma_{dyn}$, is based on a comparison of the widths $\sigma$ obtained for data and mixed background eventwise distributions respectively. We define
\begin{equation}
\sigma_{dyn}=sign(\sigma_{data}^2 - \sigma_{mix}^2) \sqrt{\mid\sigma_{data}^2 -
\sigma_{mix}^2 \mid}
\end{equation}
\noindent the quantity that is illustrated in the right hand panel of Fig.~\ref{fig:fig6}. Note that in (4) $\sigma_{data}^2$ and $\sigma_{mix}^2$ are the {\bf relative} fluctuation square
width, i.e. $\sigma_{data}^2=\sigma^2$ (distributions)/$<K/\pi>^2$, normalized to the
ensemble data average of square eventwise $K/\pi$. Thus the \% scale of $\sigma_{dyn}$. \\
\begin{figure}
\begin{center}
\includegraphics[width=7cm]{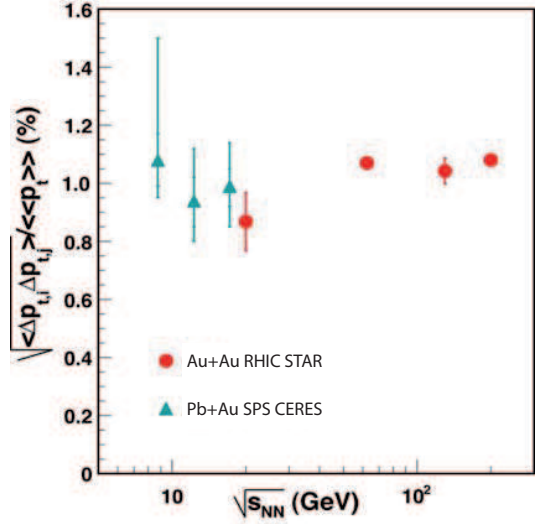}
\caption{Relative two particle $p_T$ correlator at SPS \cite{53} and RHIC \cite{52}
energies.}
\label{fig:fig16}
\end{center}
\end{figure} \\
$<K/\pi>$ represents the mean value of the data distribution shown in the left hand panel
of Fig.~\ref{fig:fig6}. We note, for clarity, that this average ratio does {\bf not} represent the
global $<K>/<\pi>$ ratio, at given $\sqrt{s}$ , which is obtained after systematic
efficiency and acceptance corrections, from a completely different ensemble analysis. On
the contrary, any event-by-event analysis is performed with the raw identified charged
particle tracks, recorded in the individual event, without any corrections. Also, we
repeat that the $K/\pi$ ratio analyzed here \cite{35} refers to both charge signs, $K^+ +
K^-/\pi^+ + \pi^-$. Is the rise toward low $\sqrt{s}$ (and thus toward $\mu_B=500 \: MeV$)
in the $K/\pi$ fluctuation an indication of non-trivial dynamics \cite{36} at high
$\mu_B$? It has been shown \cite{59,60} that $\sigma_{dyn}$ in the $K/\pi$ case where
$<K>$ is much smaller than $<\pi>$ (note that these are the mean values of multiplicity
of tracks {\bf in the acceptance} of the analysis, not the corrected multiplicities in
$4\pi$ or at $-0.5<y<+0.5$) is given approximately by
\begin{equation}
\sigma_{dyn} (K/\pi)\propto \frac{<(\Delta K)(\Delta \pi)>}{<K>}
\end{equation}
\noindent where the numerator expresses the true correlation effect but the denominator
enters a feature characterizing the observational mode. This influence may dominate the
signal as is shown in Fig.~\ref{fig:fig18}, the fall off in $<K>$ thus causing the increase of the signal toward low $\sqrt{s}$ \cite{61}. The same effect is exhibited by the STAR data
\cite{62} at $\sqrt{s}=62.4$ and $200\: GeV$ in Fig.~\ref{fig:fig19}. It shows the centrality
dependence of a related fluctuation measure (see below), with a steep, monotonous increase
toward peripheral collisions with decreasing accepted $<K>$ and $<\pi>$.

\noindent The fluctuation measure employed in the STAR analysis represents an alternative
to $\sigma_{dyn}$ employed by NA49: $\nu_{dyn}$ expresses \cite{63} the difference of the
relative event multiplicities, $\frac{K}{<K>} - \frac{\pi}{<\pi>}$. By definition the
average value of this expression is zero. The corresponding variance of this quantity is
\begin{equation}
\nu=\Big< \Big(\frac{K}{<K>} - \frac{\pi}{<\pi>}\Big)^2 \Big>=\frac{<K^2>}{<K>^2} - 2
\frac{<K \pi>}{<K><\pi>} + \frac{<\pi^2>}{<\pi>^2}.
\end{equation}

\noindent For purely statistical fluctuations (Poisson statistics) this expression is
reduced to
\begin{equation}
\nu_{stat}=\frac{1}{<K>} + \frac{1}{<\pi>}
\end{equation}
\noindent and thus we have
\begin{equation}
\nu_{dyn}=\nu - \nu_{stat}=\frac{<K(K-1)>}{<K>^2} - 2 \frac{<K\pi>}{<K><\pi>} +
\frac{<\pi(\pi-1)>}{<\pi>^2}.
\end{equation} 
\begin{figure}
\begin{center}
\includegraphics[width=7cm]{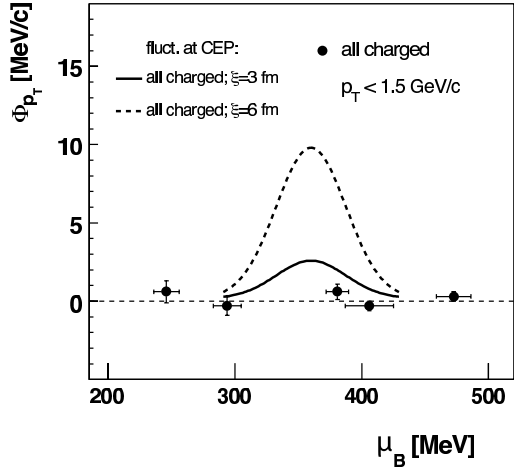}
\caption{Transverse momentum eventwise fluctuation in central $Pb+Pb$ collisions
\cite{54}; the SPS energies are represented by the corresponding values of $\mu_B$. Also illustrated are enhancements expected from a critical point \cite{26,56,57,58}.}
\label{fig:fig17}
\end{center}
\end{figure}
For further analysis of this variable see \cite{63}. It analyzes {\bf relative}
eventwise multiplicities like $K/<K>$, similar to the approach taken \cite{35} with
$\sigma_{dyn}$, and thus shares the feature to be inversely proportional to number of
kaons (and pions) involved in the eventwise analysis, i.e. the $<K>$, $<\pi>$ employed
above. The STAR results in Fig.~\ref{fig:fig19} for $\nu_{dyn}\:(K \pi)$, clearly exhibit this
property. In these minimum bias runs the geometrical reconstruction volume (the
acceptance) stays constant, of course, and unless there are systematic variations of the
reconstruction efficiency with centrality, the accepted average eventwise multiplicities
$<K>$ or $<\pi>$ above are proportional also to the overall, ''true'' average midrapidity
multiplicity $<dN/d \eta>$ for charged particles, at each centrality; this is assumed in
employing $dN/d \eta$ as the ordinate. We note that this procedure can {\bf not} be
extended to the NA49 data on energy dependence of $K/\pi$ fluctuation (as is, in fact,
done in ref. \cite{62}) as the acceptance was changed with $\sqrt{s}$ in the NA49 SPS
runs, thus $<K>$ and $<\pi>$ are {\bf not} simply proportional to the variation of
$dN/d\eta$ with energy \cite{64}, in this case. These complications are, however, taken
well into account in the fit shown in Fig.~\ref{fig:fig18}.

\noindent We conclude that the rise in the $K/\pi$ fluctuation signal represents, first of
all, a consequence of considering the {\bf relative} eventwise multiplicities and their
fluctuations, both in $\sigma_{dyn}$ and $\nu_{dyn}$. A new observable is required, to
focus exclusively on the genuine correlation effect expressed by $<(\Delta K)(\Delta
\pi)>$, or by the difference between $<K/\pi>$ and $<K>/<\pi>$. The question of signals
from the expected new physics thus remains open.\\
\begin{figure}
\begin{center}
\includegraphics[width=7cm]{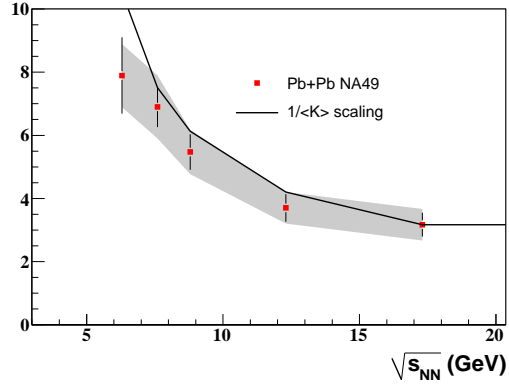}
\caption{Relative fluctuation of $K^+ +K^-/\pi^+ + \pi^-$, with $\sigma_{dyn}$ in percent,
similar as in Fig.~6. Also indicated is a dependence \cite{59,60} on the inverse number of average accepted kaons, at each $\sqrt{s}$.}
\label{fig:fig18}
\end{center}
\end{figure}

\subsection{Baryon - Strangeness Correlations: the $K/p$ Fluctuation Signal}

\noindent A further fluctuation - correlation signal, addressing the connection between
strangeness and baryon number, has recently been proposed by Koch, Majumder and Randrup
\cite{65}, as a diagnostic signal to study the QCD matter in the vicinity of the
hadronization line in the [$T, \: \mu]$ diagram of Fig.~\ref{fig:fig1}. The idea: in a medium
consisting of deconfined quarks, the correlation between strangeness and baryon number is
trivially caused by the strange quark quantum numbers. Baryon and strangeness number are strictly correlated. On the other hand, in a hadron-resonance state, strangeness is
carried both by baryon number zero meons, and by hyperons carrying strangeness and baryon number. Here, the baryon-strangeness number correlation depends on the baryonchemical potential and, thus, on $\sqrt{s}$ of $A+A$ collisions.

\noindent A fascinating proposal arises: the baryon-strangeness correlation coefficient
defined as \cite{65}
\begin{equation}
C_{BS}=-3 \: \frac{<BS>-<B><S>}{<S^2>-<S>^2}= -3 \: \frac{\sigma_{BS}}{\sigma_s^2}
\end{equation}

\noindent is formulated to be unity in a simple quark-gluon ''gas'', with deviation from
unity in a hadron-resonance system, depending on $\mu_B$. On the one hand it is related to
the quark susceptibilities, for $u, \: d, \:s$ flavours \cite{66}
\begin{equation}
C_{BS} = -3 \: \frac{\chi^{BS}_{11}}{\chi^S_2} = -3 \: \frac{\chi^{us}_1 +
\chi^{ds}_1}{\chi^{ss}_2}
\end{equation}

\noindent that have recently been obtained from a lattice calculation with three dynamical
quark flavours at finite baryochemical potential. Please note that lattice QCD calculates
susceptibilities as a function of (lattice) temperature, across the deconfinement
temperature $T_c$, such that in principle, $C_{BS}$ lattice predictions are available both
for the deconfined limit $(T >T_c)$ and for QCD matter in the hadron-resonance limit
$(T<T_c)$. Thus, both the situations concerning baryon-strangeness correlation,
anticipated in $C_{BS}$ analysis, are thus accessible for the first time. \\
\begin{figure}
\begin{center}
\includegraphics[width=7cm]{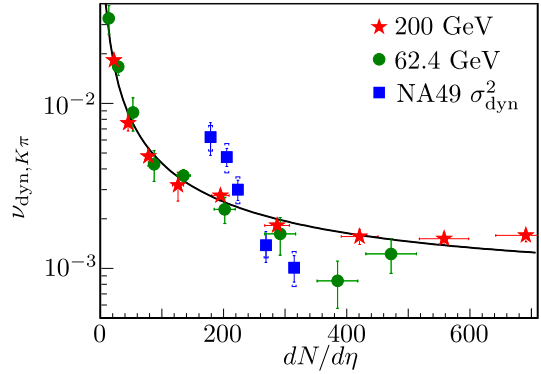}
\caption{Centrality dependence of the relative $K/\pi$ fluctuation measure $\nu_{dyn}$ in
minimum bias collisions \cite{62} of $Au+Au$ at two RHIC energies.}
\label{fig:fig19}
\end{center}
\end{figure}

\noindent Finally the $BS$ correlation can be approximatively related \cite{67} to the
experimentally observable fluctuation of the eventwise kaon to proton ratio (more
specifically: the $K^+$ to proton ratio). The $K^+$ carries the major fraction of
(anti-)strange quarks, the anti-hyperons not being accessible to any event-by-event
analysis anyhow. The proton multiplicity at high $\mu_B$ represents about one-half of the
net baryon number density, the neutron remaining not identified. A preliminary theoretical
analysis \cite{67} suggests an approximate relationship connecting the experimentally
observed $K^+$ to proton fluctuation with $C_{BS}$:
\begin{equation}
\sigma^2 (K^+/p)=\frac{<(\Delta K^+)^2>}{<K^+>^2} + \frac{<(\Delta p)^2>}{<p>^2} +
\frac{4}{3} \: \frac{C_{BS}}{<p>},
\end{equation}
\noindent thus closing the chain from lattice QCD at finite baryochemical potential, to
observable $K^+/p$ fluctuation.

\begin{figure}
\begin{center}
\includegraphics[width=7.5cm]{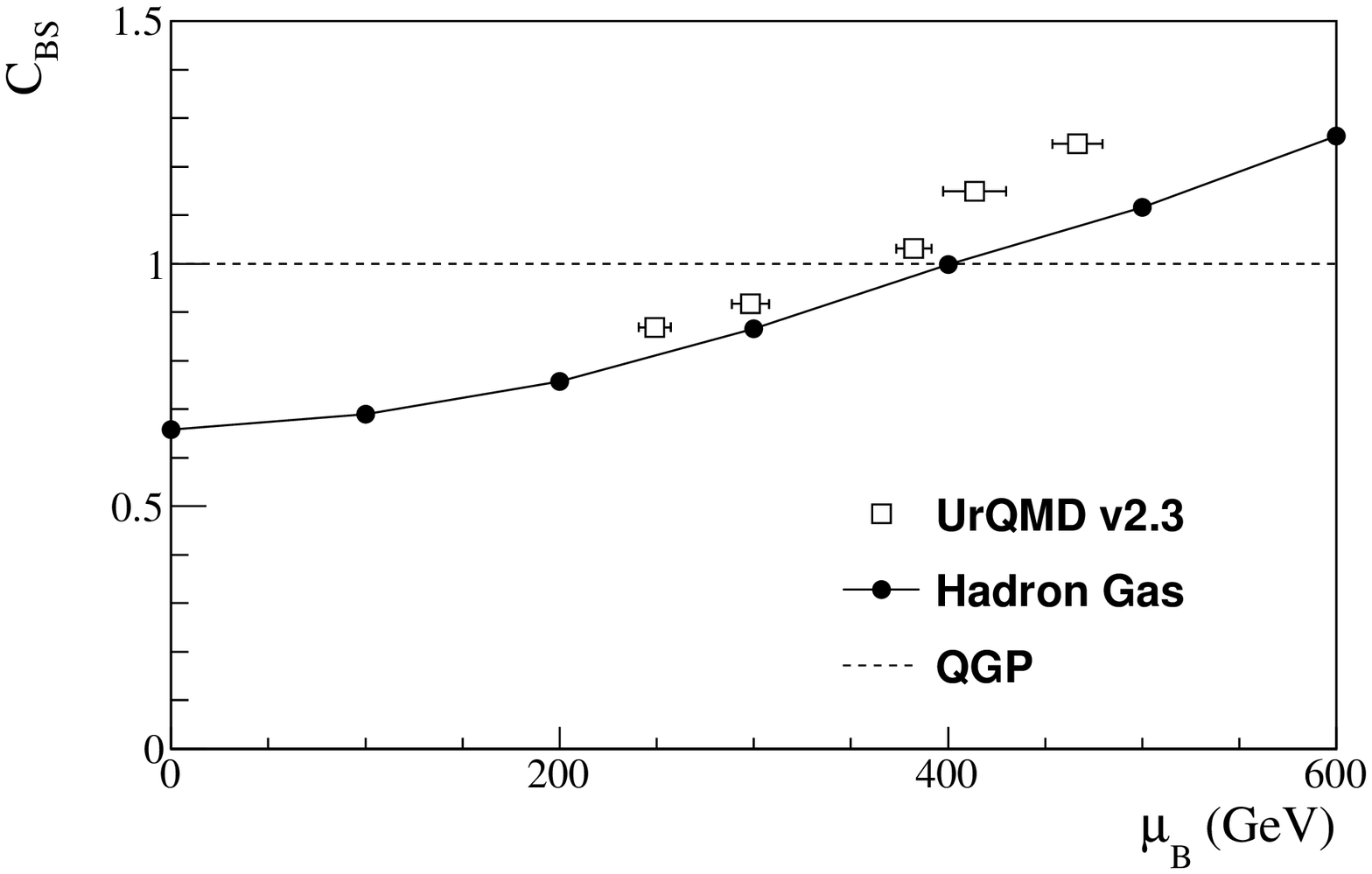}
\includegraphics[width=7.5cm]{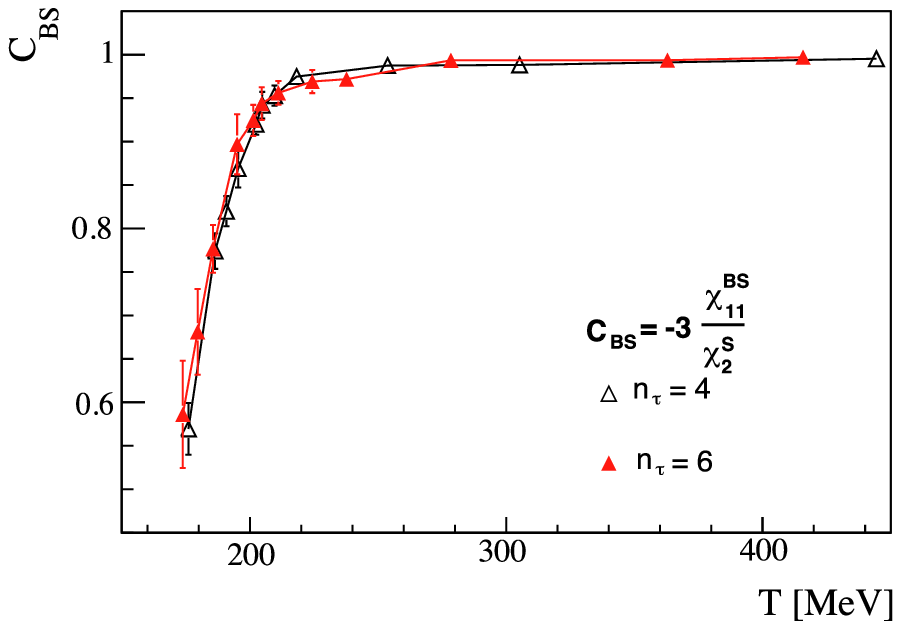}
\caption{Baryon-strangeness correlation coefficient $C_{BS}$ \cite{65}. \newline Left:
expectation for a quark plasma, a grand canonical hadron-resonance gas, and for the
transport model UrQMD \cite{61}. \newline Right: $C_{BS}$ from equ. (11) extracted from a lattice calculation \cite{27} at zero $\mu_B$.}
\label{fig:fig20}
\end{center}
\end{figure}

\begin{figure}
\begin{center}
\includegraphics[width=8cm]{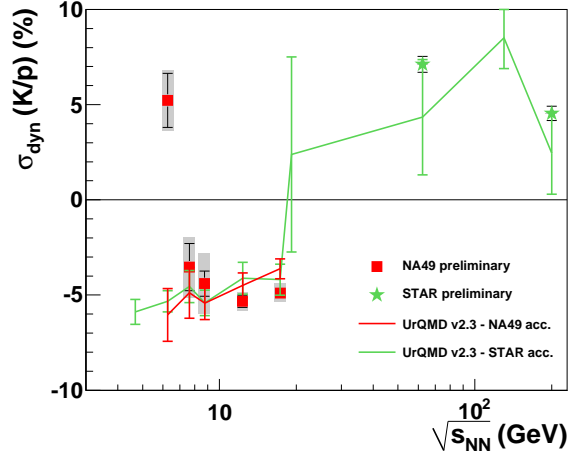}
\caption{Preliminary results for the dynamical fluctuation of the $K^+ + K^-/p+
\overline{p}$ ratio, from NA49 \cite{61} and STAR \cite{68}. Also shown is the UrQMD model
prediction.}
\label{fig:fig21}
\end{center}
\end{figure}

\noindent We show in Fig.~\ref{fig:fig20} the predictions made by Koch et al. \cite{65} for an ideal parton gas, and for a hadron-resonance gas obtained from the statistical model along the chemical freeze-out line. Also shown are corresponding UrQMD calculations \cite{61} which confirm the grand canonical predictions. Finally Fig.~\ref{fig:fig20} shows $C_{BS}$ from equ. (11) as extracted from recent 3 flavour lattice calculations \cite{27}, which indeed confirm at zero $\mu_B$ the expected saturation at unity for $T>T_c$, and the decline at $T$ well below $T_c$ where hadronic freeze-out (of course not present in lattice calculations)
would occur. Such calculations are also becoming available \cite{66} at finite $\mu_B$,
such that the connection to $K^+/p$ fluctuation data could be explored. The first data
refer to studies without kaon and proton charge separation, i.e. of ($K^+ + K^-)/p +
\overline{p}$, and are shown in Fig.~\ref{fig:fig21} along with the UrQMD simulation. The NA49 data \cite{61} at SPS energies show a negative correlation except for the lowest energy where the signal shoots up. At RHIC energy the STAR data \cite{68} give a positive correlation, and, with the notable exception of the point at $20 \: AGeV$, the UrQMD results also exhibit this change of sign. It may be understood in terms of stopping net baryon number. At CERN energy $<\overline{p}>$ is much smaller than $<p>$, and the latter are thus not newly produced protons but the ''valence'' once shifted from beam to midrapidity. An upward fluctuation of the stopping mechanisms simultaneously leads to more protons at
midrapidity and more stopped energy, available for meson production. In fact the $\pi/p$
fluctuation measured by NA49 \cite{35} at this energy also exhibits a negative
correlation. At RHIC, on the contrary, both $p$ and $\overline{p}$ at midrapidity are
newly produced , competing with kaon pair production. \\
 \\
The rise to positive fluctuation at the lowermost SPS energies remains
unexplained. The fluctuation of $K^+/p$, of key interest in the light of $C_{BS}$ and
lattice QCD analysis, will soon be available and may help to understand the low energy
behaviour in Fig.~\ref{fig:fig21}. What we are looking for is a reflection of the almost singular
behaviour of lattice susceptibilities (Fig.~\ref{fig:fig2} and~\ref{fig:fig3}) at $\mu_B \rightarrow 3 \: T_c$, for light quarks but also for $s$ quarks, to some extent.

\section{Outlook}
\noindent If it may be legitimate to write down a ''belief'' held by most if not all
investigators we might express the expectation that the crossover nature of the phase
boundary of Fig.~\ref{fig:fig1} persists at least up to the domain of $\mu_B =200 \: MeV$, i.e. to top SPS energy. Our present interest concerns the physics in the vicinity of the phase
boundary at higher $\mu_B$ e.g.
\begin{itemize}
\item a critical point of QCD,
\item an adjacent first order nature of phase transition,
\item a related '' softest point'' of the equation of state, governing eliptic flow,
\item a disentangling of the QCD chiral transition, from the confinement transition,
\item the related chiral restoration mechanism,
\item the exact position in the $[T, \: \mu_B]$ plane of all this,
\item the existence, and domain, of a further QCD phase, and phase boundary, at high
$\mu_B$ (''quarkyonium'').
\end{itemize} 

\noindent As a consequence of the expected need for substantial work at relatively low
energies, say $\sqrt{s}\le 20\: GeV$, there arises an unfortunate split in the RHIC
community, amusingly just coinciding with either accelerating or decelerating after RHIC
injection from the AGS. It thus appears  difficult to agree on an extended low energy scan
at RHIC but this is exactly what is required. With the possible exception of top SPS
energy, $\sqrt{s}=17.3 \: GeV$, the statistics and systematics of the 2000-2002 SPS low
energy runs in the domain $250<\mu_B<450 \: MeV$ - which many consider the potentially
most interesting - was far from adequate. Furthermore, the recent advances of lattice
theory, both toward realistic quark masses and 3 flavour calculations, and toward higher
statistical moments, and finite $\mu_B$, lead one to expect that all the topics enumerated
above will come within reach of rigorous theory. Of course, at this point we have to
acknowledge, also, the present dichotomy among lattice QCD groups, about virtually
everything - but then how about some experimental progress? For example an interesting
physics topic, to be tackled next, consists in higher moments and cumulants of fluctuating
quantities, such as quark densities, which exhibit interesting structures \cite{27} in the
vicinity of the critical (transition) temperature. These quantities are referring to the
higher moments (beyond the width) of event by event fluctuating quantities, i.e. to
symmetric (curtosis) or asymmetric (skewness) deviations from the normal distribution
\cite{27}. For example, Stephanov \cite{69} has shown that such higher moments can better
constrain the range $\xi$ of the critical domain corresponding to a potential critical
point (c.f. Fig.~\ref{fig:fig17}). First experimental data for the curtosis have been shown by STAR \cite{70}, for fluctuation of conserved charges at mid-rapidity. The implication of
conservation laws - the fact that a single $A+A$ event in its totality represents a
''grand microcanonical'' ensemble - has been discussed in \cite{71}. These considerations
should give rise to a new class of physics observables.

\end{document}